\documentclass[showpacs,aps,amsmath,amssymb,twocolumn]{revtex4}
\usepackage{graphicx}
\usepackage{dcolumn}
\usepackage{bm}

\begin{document}

\title{Self-interactions in the space-time of a scalar-tensor cosmic string}

\author{I. V. L. Costa and F. A. Oliveira}
\affiliation{Instituto de F\'{\i}sica, Universidade de Bras\'{\i}lia,\\
CEP: 70910-900, Bras\'{\i}lia, DF, Brazil}

\author{M. E. X. Guimar\~aes}
\affiliation{Departamento de Matem\'atica, Universidade de Bras\'{\i}lia\\
CEP: 70910-900, Bras\'{\i}lia, DF, Brazil}

\author{Fernando Moraes}
\affiliation{Departamento de F\'{\i}sica,  Universidade Federal da Para\'{\i}ba,
58051-970 Jo\~ao Pessoa, PB, Brazil}

\begin{abstract}We study the effect of the geometry and topology of a
scalar-tensor cosmic string space-time on the electric and
magnetic fields of line sources. It is shown that the dilatonic
coupling of the gravity induces effects along the string
comparable to a current flow allowing for forbidden regions near
the string.

\end{abstract}

\keywords{Scalar-tensor gravities, classical field theory in curved
spaces}
\pacs{98.80.Cq;11.27.+d}
\maketitle

\section{Introduction}

It is a well known fact~\cite{witt,vilenkin} that a point charge
in a static gravitational field experiences an electrostatic force
due to the deformation of its electric field lines produced by the
geometry of space-time. The net effect is a self-force on the
charge. The study of self-forces on electric charges in the
presence of topological defects started when Linet~\cite{linet}
and Smith~\cite{smith} found the electrical self-force on a point
charge in the presence of a cosmic string. Bezerra de Mello {\it
et al.}~\cite{bezerra} found both the electric self-force on a
line of charge and the magnetic self-force on an electrical
current parallel to a cosmic string.

All the above-mentioned results have been obtained in the
framework of Einstein's gravity. Here, we intend to generalize
these results considering a scalar-tensor gravity. The motivation
for this relies on the fact that, theoretically, the possibility
that gravity might not be fundamentally Einsteinian is gathering
credence. This is in part a consequence of superstring
theory~\cite{green}, which is consistent in ten dimensions (or
M-theory in eleven dimensions), but also the more phenomenological
recent developments of ``braneworld'' scenarios~\cite{ADD,RS} have
motivated the study of other gravitational theories in
four-dimensions. In fact, the origin of the (gravitational) scalar
field can be many: the scalar field arising from the size of the
compactified internal space in the Kaluza-Klein theory; the zero
mode (dilaton field) described by a symmetric second-rank tensor
behaving as space-time metric at low energy level in the closed
string theory; the scalar field in a brane world scenario; and
more~\cite{maeda}. In any case, clearly, if gravity is essentially
scalar-tensorial there will be direct implications on observed
effects both in the small scale scenarios of alternative theories
of gravity~\cite{KKC,BHC,CRS, SNN} and in the large scale
cosmological scenarios from modified gravity~\cite{MNG}.

It is our interest here to consider small-scale effects and to
show these modifications to the case of a scalar-tensor cosmic
string. In this work we find the self-energy and self-force on a
line source in the presence of a scalar-tensor cosmic string. We
look both at the case of the source being a line with uniform
charge density or a constant current and, in  doing this, we get
information of both the electric and magnetic self-interactions.
We have chosen to apply here the method developed by Grats and
Garcia in the ref. \cite{grats} which consists in describing the
renormalised quantities using Riemannian coordinates, and in
choosing the origin of the Riemannian frame at the centre of the
geodesics. This method has been proved to be quite successful and,
with our calculations, we generalize the work previously done by
one of us in  ref. \cite{bezerra}.

The layout of this work is the following. In  section 2, we
calculate the self-energy and the self-force  on a line source in the presence of a scalar-tensor cosmic string. We compare these expressions with the ones obtained
in the General Relativity case. In the section 3, we summarize our
main results.

\section{The self-energy and the self-force on a line source}

The solution of Einstein equations for a straight infinite cosmic
string in the framework of the scalar-tensor theory was found, in
the weak field approximation, by one of us ~\cite{guimaraes}. The
line element, in cylindrical coordinates, with the string placed
along the \textit{z}-axis, is given by
\begin{eqnarray}
\label{1} ds^{2} & = & [1+8G_{0}\mu\alpha^{2}(\phi_{0})\ln\left(
R/R_{c}\right)]  \nonumber \\
& & \cdot\left(-dt^{2}+dz^{2}+dR^{2}+\beta^{2}R^{2}d\varphi^{2}\right), 
\end{eqnarray}
where $\beta^{2}=(1-8G_{0}\mu)$. Here, $G_{0}$ is the
$\phi_{0}$-dependent effective gravitational constant defined as
$G_{0}\equiv G_{*}A^{2} (\phi_{0})$ ($G_{*}$ being some bare
gravitational coupling constant, $A^{2}$ is an arbitrary function
of the dilaton field), $\alpha^{2}(\phi_{0})$ is a Post-Newtonian
parameter which expresses the coupling between matter and the
dilaton field and $\phi_0$ is the value of the dilaton field
evaluated by solar systems experiments~\cite{damour}.

Since our problem is effectively two-dimensional, we will work
with the (\textit{t}=const, \textit{z}=const) section of metric
(\ref{1}). For convenience, we perform the change of variables
\begin{equation}
\label{2} R=\frac{r^{\beta}}{\beta}
\end{equation}
in order to write the two-dimensional metric in conformal fashion,
\begin{eqnarray}
\label{3}
ds^{2}=\exp\left[-\Omega(r)\right]\left(dr^{2}+r^{2}d\varphi^{2}\right),
\end{eqnarray}
where
\begin{equation}
\label{4}
\Omega(r)=-\ln\left\{r^{2\beta-2}\left[1+8G_{0}\mu\alpha^{2}(\phi_{0})\ln\left
(\frac{r^{\beta}}{\beta R_{c}}\right)\right]\right\}.
\end{equation}

Consider now an infinitely long straight wire, placed parallel to
the \textit{z}-axis, at the position $\vec{r}\,'=(r',\varphi')$,
with uniform linear charge density $\lambda$. The corresponding
three-dimensional charge density is
\begin{equation}
\label{5}
\rho(\vec{r})=\lambda\frac{\delta(r-r')\delta(\varphi-\varphi')}{r}=
\lambda\delta^{(2)}(\vec{r}-\vec{r}\,').
\end{equation}

Given an arbitrary distribution of charge $\rho(\vec{r})$,the
electrostatic energy is given by~\cite{jackson}
\begin{equation}
\label{6} W=\frac{1}{2} \int\int
\rho(\vec{r})G^{(3)}(\vec{r},\vec{r}\,')\rho(\vec{r}\,') d^{3}r
d^{3}r',
\end{equation}
where $G^{(3)}(\vec{r},\vec{r}\,')$ is the three-dimensional
Laplacian Green function. In our case, the problem is effectively
two-dimensional. So, the electrostatic energy per unit length is
\begin{equation}
\label{7} \frac{U_{ele}}{l}=\frac{1}{2} \int\int
\rho(\vec{r})G_{d}^{(2)}(\vec{r},\vec{r}\,')\rho(\vec{r}\,')
d^{2}r d^{2}r', 
\end{equation}
where $G_{d}^{(2)}(\vec{r},\vec{r}\,')$ is the two-dimensional
Green function in the space described by metric (\ref{3}).

By substituting (\ref{5}) in (\ref{7}) and taking the limit
$\vec{r}\rightarrow\vec{r}\,'$ the electrostatic energy per unit
length is obtained as
\begin{equation}
\label{8}
\frac{U_{ele}}{l}=\frac{\lambda^{2}}{2}G_{d}^{(2)}(\vec{r}\,',\vec{r}\,')|_{reg},
\end{equation}
where $G_{d}^{(2)}(\vec{r}\,',\vec{r}\,')|_{reg}$ is the
regularized Green function
\begin{equation}
\label{9}
G_{d}^{(2)}(\vec{r}\,',\vec{r}\,')|_{reg}=\lim_{\vec{r}\rightarrow\vec{r}\,'}
\left[G_{d}^{(2)}(\vec{r},\vec{r}\,')-G^{(2)}(\vec{r},\vec{r}\,')\right].
\end{equation}
Here, $G^{(2)}(\vec{r},\vec{r}\,')$ is the Euclidean Green
function, solution of the two-dimensional Poisson equation
\begin{equation}
\label{10}
\Delta_{E}G^{(2)}(\vec{r},\vec{r}\,')=-4\pi\delta^{2}(\vec{r}-\vec{r}\,'),
\end{equation}
where $\Delta_{E}$ is the two-dimensional Euclidean Laplacian.

The regularization, by extraction of the divergent part of the
Green function, is necessary in order to obtain a finite result
for the energy. On the other hand, this also guarantees that no
finite self-energy survives in the absence of the cosmic string.

It has been shown by Grats and Garcia~\cite{grats} that, since any
two-dimensional metric can be put in the form given by eq.
(\ref{3}), the Laplacian Green function in the corresponding space
can be expanded in terms of the geodesic distance
$\sigma(\vec{r},\vec{r}\,')$, such that
\begin{eqnarray}
\label{11}
-G_{d}^{(2)}(\vec{r},\vec{r}\,') & = & \ln\left[2\sigma(\vec{r},\vec{r}\,')\right]
+\Omega(\vec{r}) \nonumber \\
& & +(2\sigma)\frac{t^{a}t^{b}\theta_{ab}}{24}+O(\sigma^{2}),
\end{eqnarray}
where $2\sigma(\vec{r},\vec{r}\,')$ is the squared geodesic
distance between the points $\vec{r}$ and $\vec{r}\,'$, $t^{a}$ is
the tangent vector to the geodesic at point $\vec{r}$, and the
tensor $\theta _{ab}$ has the form
\begin{equation}
\theta _{ab}=\nabla _a\Omega \nabla _b\Omega -\frac 12g_{ab}\nabla
_c\Omega \nabla ^c\Omega +\nabla _{ab}^2\Omega ,
\end{equation}
where $\nabla_a$ is the covariant derivative  and $g_{ab}$ is the
metric tensor.

Since the Euclidean Green function can be written as
\begin{equation}
\label{12}
G^{(2)}(\vec{r},\vec{r}\,')=-\ln\left[2\sigma(\vec{r},\vec{r}\,')\right],
\end{equation}
it turns out that ~\cite{grats}
\begin{equation}
\label{13}
G_{d}^{(2)}(\vec{r}\,',\vec{r}\,')|_{reg}=-\Omega(\vec{r}\,').
\end{equation}
This equation shows explicitly that the self-energy appears from
the geometry induced by the defect.

Using (\ref{13}), (\ref{8}), (\ref{4}) and returning to the
variable $R$, eq. (\ref{2}), we get
\begin{eqnarray}
\label{U} \frac{U_{ele}}{l}  & = & \frac{\lambda^{2}}{2} \ln \left\{
(\beta R)^{2-2/\beta} \right. \nonumber \\
& & \cdot \left. \left[1  + 8G_{0}\mu\alpha^{2}(\phi_{0})
\ln\left(\frac{R}{R_{c}}\right)\right] \right\}
\end{eqnarray}
and
\begin{eqnarray}
\label{13.5}
& & \frac{\vec{F}_{ele}}{l}  =  -\vec{\nabla}\left(\frac{U_{ele}}{l}\right)=\frac{\lambda^{2}}{2R} \nonumber \\
&  & \cdot \left[2\left(\frac{1}{\beta}-1\right)  -\frac{8G_{0}\mu\alpha^{2}(\phi_{0})}{1+8G_{0}\mu\alpha^{2}(\phi_{0})
\ln\left(\frac{R}{R_{c}}\right)}\right]\widehat{r}.
\end{eqnarray}
Notice that, when $\alpha^{2}(\phi_{0})\rightarrow 0$, we recover
the Einstein gravity result, equation (18), of reference
\cite{bezerra}.

Now, in order to study the magnetic self-force, we consider the infinitely long straight wire carrying the current density
\begin{equation}
\vec{J}(\vec{r})=I\frac{\delta(r-r')\delta(\varphi-\varphi')}{r}\hat{z}=
I\delta^{(2)}(\vec{r}-\vec{r}\,'){r}\hat{z}.
\end{equation}
The magnetic analogue of equation (\ref{7}) is \cite{jackson}
\begin{equation}
\frac{U_{mag}}{l}=\frac{1}{2c^2} \int\int
\vec{J}(\vec{r})\cdot \vec{J}(\vec{r}\,') G_{d}^{(2)}(\vec{r},\vec{r}\,')
d^{2}r d^{2}r', 
\end{equation}
which results in
\begin{equation}
\frac{U_{mag}}{l}=\frac{I^{2}}{2c^2}G_{d}^{(2)}(\vec{r}\,',\vec{r}\,')|_{reg}, \label{umag}
\end{equation}
analogous to equation (\ref{8}).

As shown in \cite{bezerra}, the magnetic self-force per unit length is given by
\begin{equation}
\frac{\vec{F}_{mag}}{l}  =  \vec{\nabla}\left(\frac{U_{mag}}{l}\right). \label{fmag}
\end{equation}
Equations (\ref{fmag}), (\ref{umag}) combined with (\ref{13}) and (\ref{4}) result in
\begin{eqnarray}
& & \frac{\vec{F}_{mag}}{l}=-\frac{I^{2}}{2c^{2}
R} \nonumber \\
& & \cdot \left[2\left(\frac{1}{\beta}-1\right)
-\frac{8G_{0}\mu\alpha^{2}(\phi_{0})}{1+8G_{0}\mu\alpha^{2}(\phi_{0})\ln\left(\frac{R}
{R_{c}}\right)}\right]\widehat{r}. \label{fmag2}
\end{eqnarray}

Comparing the expression for the magnetic self-force, eq.
(\ref{fmag2}),  with the one for the electric self-force, eq.
(\ref{13.5}), we see that they are proportional to each other:
\begin{equation}
\frac{\vec{F}_{mag}}{l}=-\frac{I^{2}}{c^{2}\lambda^{2}}\frac{\vec{F}_{ele}}{l}.\label{prop}
\end{equation}

Let us discuss now the behavior of the self-force. From equation (\ref{13.5}) it is clear that $R$ can not be less than
\begin{equation}
\label{20}
R_{min}=R_{c}\exp\left(-\frac{1}{8G_{0}\mu\alpha^{2}(\phi_{0})}\right).
\end{equation}
In fact, writing (\ref{13.5}) in terms of $R_{min}$, we get
\begin{equation}
\frac{\vec{F}_{ele}}{l}  =  \frac{\lambda^{2}}{2R} 
 \left[2\left(\frac{1}{\beta}-1\right)  -\frac{1}{\ln\left(\frac{R}{R_{min}}\right)}\right]\widehat{r}. \label{FRmin}
\end{equation}
This is in complete agreement with the metric (\ref{1}) which makes sense only for $R>R_{min}$, as can be easily verified. 

Notice that, for $R>>R_{min}$ the behavior of both the electric and magnetic self-forces approaches that of the corresponding General relativity (GR) results \cite{bezerra}:
\begin{eqnarray}
\frac{\vec{F}_{ele}}{l} & = & \frac{\lambda^{2}}{R} 
\left(\frac{1}{\beta}-1\right) \\
\frac{\vec{F}_{mag}}{l} & = & -\frac{I^{2}}{c^{2}R} 
\left(\frac{1}{\beta}-1\right).
\end{eqnarray}
This behavior is seen in Figure 1 for $x=R/R_{min}>>1$.

\begin{figure}[h]
\centering
\includegraphics[width=2in,angle=-90]{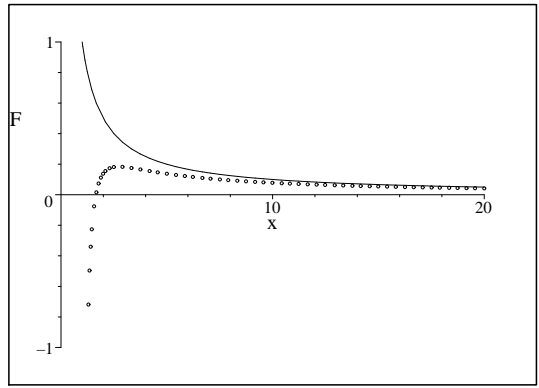}
\caption{The electric self-force per unit length $F$ as a function of $x=R/R_{min}$ for $\beta=0.5$ and $\lambda^{2}=1$. The continuous line represents the GR result.}
\label{fig1}
\end{figure}

Note in Figure 1 that the force changes its sign at some $R=R_{0}$ which can be obtained from eq. (\ref{FRmin}) with
the condition $\vec{F}_{ele}/l=0$,
\begin{eqnarray}
\label{21}
R_{0}  =  R_{min}\exp\left[\frac{\beta}{2(1-\beta)}\right].
\end{eqnarray}

Differently from the GR result which is always repulsive, the Scalar-Tensor (ST)  electric force  is attractive in the range $R_{min}<R<R_{0}$. Eq. (\ref{prop}) implies a similar analysis to $\frac{\vec{F}_{mag}}{l}$, with inverted behavior: the magnetic force is attractive for $R>R_{0}$ and repulsive for $R_{min}<R<R_{0}$. 

\section{Conclusions}

In  ref. \cite{peter}, one of us investigated the internal
nature of ordinary cosmic vortices in some scalar-tensor
extensions of gravity. Solutions were found for which the dilaton
field condenses inside the vortex core. These solutions could be
interpreted as raising the degeneracy between the eigenvalues of
the effective stress-energy tensor, namely the energy per unit
length $U$ and the tension $T$, by picking a privileged spacelike
or timelike coordinate direction; in the latter case, a {\it phase
frequency threshold} occured that is similar to what is found in
ordinary neutral current-carrying cosmic strings. It has been
found that the dilaton contribution for the equation of state,
once averaged along the string worldsheet, vanishes, leading to an
effective Nambu-Goto behavior of such a string network in
cosmology, {\it i.e.} on very large scales. It has been shown also that,
on small scales, the energy per unit length and tension depend on
the string internal coordinates in such a way as to permit the
existence of centrifugally supported equilibrium configuration,
also known as {\it vortons}, whose stability, depending on the
very short distance (unknown) physics, can lead to catastrophic
consequences on the evolution of the Universe.

This statement is in a remarkable agreement with our present
result. Indeed, in this work we determined the electric and
magnetic self-interaction of line sources in the presence of a
scalar-tensor cosmic string. When the source is placed away from
the defect (long scales), the self-interaction is quite similar to
the result obtained in the General Relativity framework. On the
other hand, near the defect, the dilaton field introduces
interesting new phenomena: a forbidden region $0<R<R_{min}$ and an
inversion of the electric (magnetic) self-force behavior, going
from repulsive (attractive) in the far region to attractive
(repulsive) in the near region.

Our main conclusion is that the coupling between a cosmic string
and a scalar-tensor gravity is shown to induce effects along the
string comparable to a current flow in the sense that the
resulting effective stress-energy tensor eigenvalues, the energy
per unit length $U$ and tension $T$, were no longer degenerate,
due to the presence of the dilaton. Our model reproduces an
Abrikosov flux tube coupled to a disclination in  a condensed
matter system \cite{fur}. We plan to continue our future works in
this direction.

\acknowledgments{This work was partially supported by PRONEX/FAPESQ-PB, CNPq and
CAPES (PROCAD)}.

\end{document}